\documentclass[reprint,superscriptaddress,amsmath,amssymb,aps,prb]{revtex4-2}

\usepackage{graphicx}
\usepackage{dcolumn}
\usepackage{bm}
\usepackage{tabularx}
\usepackage{hyperref}

\newcommand{\rev}[1]{#1}

\begin{document}

\title{Manipulating Topological Valley Modes in Plasmonic Metasurfaces}

\author{Matthew Proctor}
    \email{matthew.proctor12@imperial.ac.uk}
    \affiliation{Department of Mathematics, Imperial College London, London SW7 2AZ, UK}

\author{Paloma A. Huidobro}
    \affiliation{Instituto de Telecomunica{\c c}{\~o}es, Insituto Superior Tecnico-University of Lisbon, Avenida Rovisco Pais 1,1049-001 Lisboa, Portugal}
    
\author{Stefan A. Maier}
    \affiliation{Chair in Hybrid Nanosystems, Nanoinstitut M{\"u}nchen, Faculty of Physics, Ludwig-Maximilians-Universit{\"a}t M{\"u}nchen, 80539, M{\"u}nchen, Germany}
    \affiliation{Department of Physics, Imperial College London, London, SW7 2AZ, UK}

\author{Richard V. Craster}
    \affiliation{Department of Mathematics, Imperial College London, London SW7 2AZ, UK}
 
\author{Mehul P. Makwana}
    \affiliation{Department of Mathematics, Imperial College London, London SW7 2AZ, UK}
    \affiliation{Multiwave Technologies AG, 3 Chemin du Pr\^{e} Fleuri, 1228, Geneva, Switzerland}

\date{\today}

\begin{abstract}
The coupled light-matter modes supported by plasmonic metasurfaces can be combined with topological principles to yield subwavelength topological valley states of light. We give a systematic presentation of the topological valley states available for lattices of metallic nanoparticles: All possible lattices with hexagonal symmetry are considered, as well as valley states emerging on a square lattice. Several unique effects which have yet to be explored in plasmonics are identified, such as robust guiding, filtering and splitting of modes, as well as dual-band effects. We demonstrate these by means of scattering computations based on the coupled dipole method that encompass the full electromagnetic interactions between nanoparticles.
\end{abstract}

\maketitle

Plasmonics offers a unique platform for controlling light on the nanoscale \cite{maier2007plasmonics}. Coherent electron oscillations known as localised surface plasmons arise when light interacts with metallic nanostructures. These localised excitations confine light beyond the diffraction limit and have optical properties tunable by the size, material and shape of the nanostructures that host them, as well as the surrounding environment \cite{giannini2011plasmonic}. This has resulted in a plethora of applications including the waveguiding of light on the nanoscale \cite{barnes2003surface} and the exploration of chiral optical interactions with surface plasmon polaritons \cite{bliokh2015spin}. Notably, owing to advances in nanofabrication, unexpectedly long propagation lengths of surface plasmons in waveguides have been reported \cite{kress2015wedge}. 

Plasmonic metasurfaces, which are collections of plasmonic nanoparticles (NPs) in a plane arranged in different lattices, are the subject of current intense research due to their enriched properties leveraged by the flexibility in geometrical designs \cite{meinzer2014plasmonic}. 
Exploiting the radiative coupling regime between the NPs \cite{wang2018rich}, whose interactions are mediated by diffractive modes in the plane of the array (surface lattice resonances), has enabled the experimental demonstration of lasing \cite{zhou2013lasing,guo2019lasing} and Bose-Einstein condensation \cite{hakala2018bose} in plasmonic lattices, due to their dramatic quality factor enhancement. 
On the other hand, the near field coupling regime, where the distance between the NPs is very subwavelength, is currently regaining interest due to the possibility of realising topological phases of light confined at nanoscale dimensions using metal NPs as well as other nanoresonators such as dielectric NPs \cite{ling2015topological,downing2017topological,pocock2018topological,kruk2019nonlinear,wu2019dynamic, proctor2019exciting,slobozhanyuk2019near, sinev2015mapping, kruk2017edge}. This has been sparked by the potential of topological protection to provide robust light propagation immune to certain kinds of disorder and imperfections in samples; in analogy to the effects present in topological insulators, materials which are insulating in the bulk and possess protected conduction states along their edge \cite{hasan2010colloquium}. 

Topological insulators have unidirectional edge states where backscattering is entirely suppressed in the absence of magnetic impurities as there is no backwards propagating mode to couple into. These states are reliant on the fractional spin of fermions. Therefore, whilst these electronic systems have inspired research into bosonic analogues, they require alternative strategies for creating topological states \cite{ozawa2019topological, rider2019perspective, de2019engineering}. Unidirectional, i.e., non-reciprocal, edge states can be obtained for photons by breaking time-reversal symmetry through the use of strong magnetic fields \cite{wang2009observation,jin2017infrared}. However, there are constraints in miniaturising this approach to nanoscale set-ups and there is a need for alternative designs. 

Topological valley (Hall) modes are solely reliant upon energy extrema in reciprocal space, and they form a specific subclass not contingent upon particles with fractional spin or a time-reversal breaking component \cite{qian2018topology}. Whilst they are not completely protected against backscattering, the existence of a local, valley-dependent topological invariant means these states do inherit some aspects of topological protection \cite{qian2018topology}. An advantage of these topological valley modes is that they only require the breaking of an inversion symmetry and/or a mirror symmetry; consequently, topological valley states in plasmonics enable more robust plasmonic modes whilst remaining within reach of current experimental set-ups. Existing work on topological valley effects in photonics has focused on hexagonal and triangular structures \cite{gao2018topologically,gao2017valley,dong2017valley,ma2016all, chen2017valley}. In plasmonics, graphene was proposed to host topological valley modes at infrared frequencies by imposing a triangular structured doping landscape on a graphene sheet using a metagate \cite{jung2018midinfrared}, and designer (spoof) plasmons were used to experimentally probe these effects at microwave frequencies \cite{wu2017direct}. 

\begin{figure}
    \centering
    \includegraphics[width=\columnwidth]{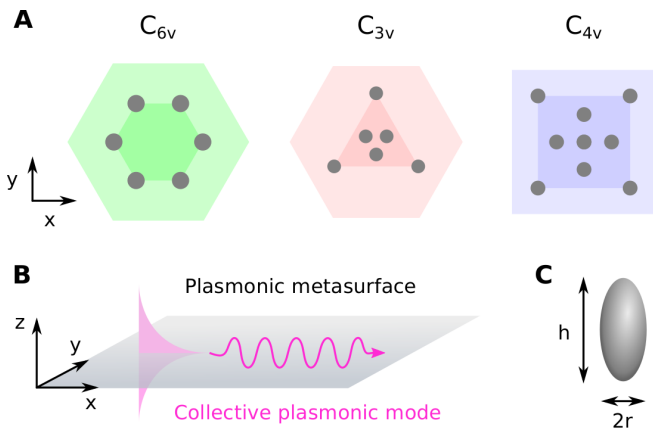}
    \caption{(A) Unit cells with $C_{6v}$, $C_{3v}$, $C_{4v}$ symmetries. The shading indicates shapes with equivalent symmetries, (B) A plasmonic metasurface: Collective plasmonic resonances arise from the coupling between localised surface plasmons on each NP; they propagate in the 2D metasurface and decay exponentially perpendicular to the plane of the metasurface, (C) Spheroidal NPs, with in-plane radius $r$ and height $h$, host a dipole moment out of the plane, $p_z$, decoupled from the in-plane dipole moment.}
    \label{fig:layout}
\end{figure}

In this paper, we show that a two-dimensional (2D) lattice of metallic NPs provides a versatile platform for demonstrating the existence of topological valley  states in plasmonics. We discuss the symmetry requirements for these states and present a systematic description of the topological valley states available. Existing cases in the literature have utilised hexagonal ($C_{6v}$) and triangular ($C_{3v}$) symmetry-induced Dirac cones and they have each respectively explored a single edge configuration. Here we classify all the possible edge configurations with these symmetries \cite{makwana2018designing, makwana2018geometrically}. Additionally, we present a square lattice with $C_{4v}$ in the Supplementary Material. 
From the properties of the edge states we design networks of regions, composed of the various symmetry types, to engineer advanced waveguiding techniques such as filtering, evanescent coupling and splitting at subwavelength scales.

\section{Methods}

We study 2D lattices of spheroidal, silver NPs (\autoref{fig:layout}C), utilising the coupled dipole method (CDM). The permittivity of the NPs is described with the Drude model,
\begin{align}
    \rev{\epsilon(\omega) = \epsilon_\infty - \frac{\omega_p^2}{\omega^2 + i \omega\gamma}}
\end{align}
with $\epsilon_\infty = 5$, $\omega_p = 8.9$~eV and $\gamma=38$~meV \cite{yang2015optical}. For each lattice arrangement we assume that the NP in-plane radius $r$, \rev{height $h$}, nearest neighbour spacing $R$, and lattice constant $a_0$ are all subwavelength. We ensure that $R > 3r$ which allows the NPs to be treated as point dipoles, as higher order resonances can be neglected \cite{weber2004propagation}. The interaction between multiple NPs is described by the Green's function, (see Supplementary Material). The in-plane and out-of-plane polarisations are orthogonal \rev{and become well separated in frequency due to the spheroidal NPs, with in-plane modes shifting to higher frequencies and out-of-plane-modes shifting to lower frequencies. This means} they can be investigated separately \cite{wang2016existence}, and here we only consider the out-of-plane polarised modes corresponding to dipole moments perpendicular to the $xy$-plane of the metasurface, \autoref{fig:layout}B. The long range terms in the Green's function are responsible for retarded interactions, which can have remarkable effects in plasmonics \cite{weber2004propagation} and in particular in topological plasmonics, where bulk-edge correspondence can break down due to retardation \cite{pocock2019bulk}. However, in this work we limit ourselves to very subwavelength arrays of NPs where near field interactions dominate. 

\section{Designing Topological Valley States}

\subsection{Engineering Dirac cones}\label{sec:dirac_cones}

\begin{figure}
    \centering
    \includegraphics[width=\columnwidth]{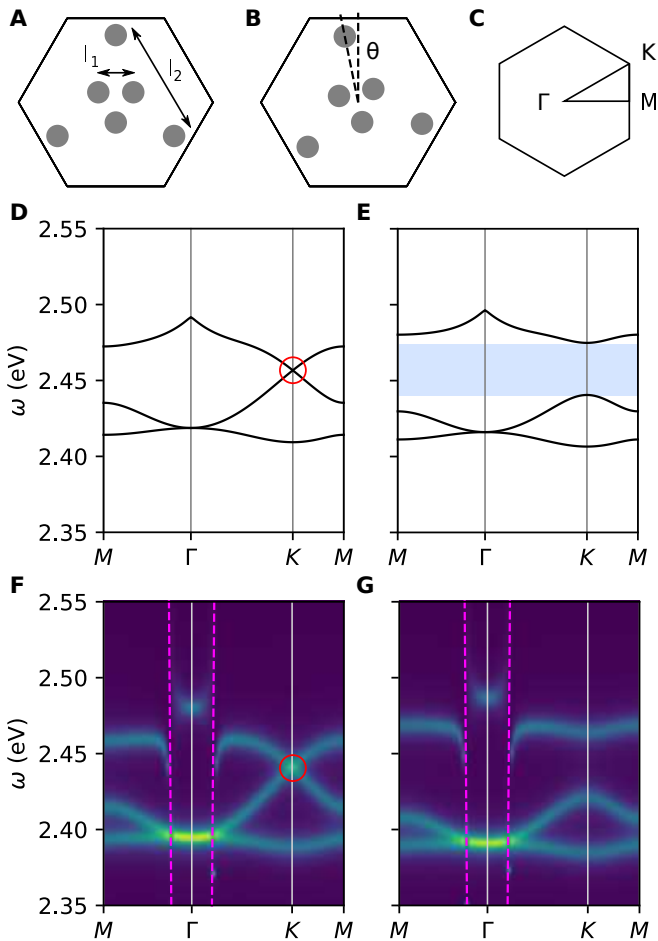}
    \caption{$C_{3v}$ case: (A) Unit cell with NP radius $r=5$~nm \rev{and height $h=30$~nm}, internal spacing $l_1 = \rev{10\sqrt{3}}$~nm, external spacing $l_2 = \rev{30\sqrt{3}}$~nm, lattice constant $a_0 = 75$~nm, (B) A $\theta=\rev{0.1}$~rad rotation removes $\sigma_v$, (C) BZ, (D) Band structure, pre-perturbation, (E) Post-perturbation, (F,G) Spectral function: The higher energy bands experience polariton-like splitting at the light line (magenta). Drude losses $\gamma = \rev{10}$~meV.}
    \label{fig:c3v_bulk}
\end{figure}
% mention somewhere that 'maximum' allowed perturbation is system dependent
% i.e. graphene metagate paper: even with max rotation, curvature around valleys is maintained

In designing topological valley states, we follow a process of engineering a Dirac cone and then breaking it. Dirac cones are degeneracies in the band structure of a system which disperse linearly and can be described by an effective Dirac Hamiltonian. This conical dispersion leads to a number of interesting transport phenomena \cite{lu2014dirac}.  Topological valley states rely on the existence of valleys, energy extrema around a lifted Dirac degeneracy, in the band structure. A large separation of two valleys in the Brillouin zone (BZ) allows a local, valley-dependent topological invariant to be defined: the valley-Chern number \cite{zhu2018design}. This quantity is calculated from the Berry curvature, which is strongly localised at the valleys provided the perturbation which breaks the Dirac cone is small enough \cite{qian2018topology}. A large perturbation results in the Berry curvature becoming less localised and the topological valley effect is diminished. 

Despite being locally defined, the valley-Chern number has been shown to guarantee the existence of edge states for a host of systems \cite{ma2017scattering}, as a result of the bulk boundary correspondence \cite{qian2018topology}; the properties of these edge states are discussed in \autoref{sec:edge_states}.

\begin{figure}
    \centering
    \includegraphics[width=\columnwidth]{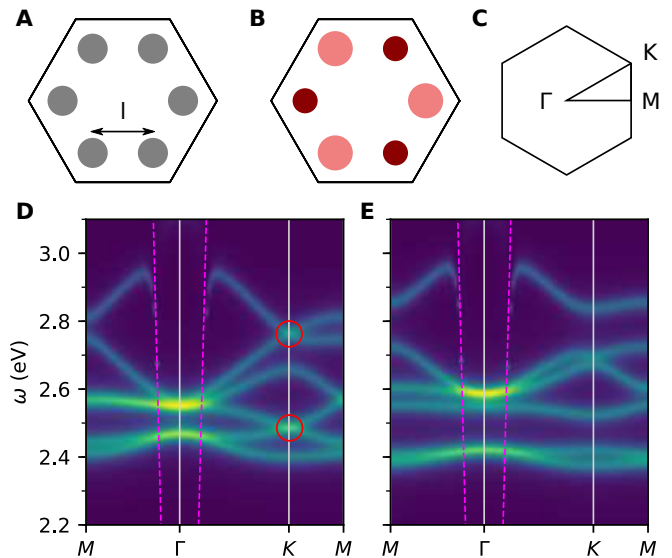}
    \caption{\rev{$C_{6v}$ case: (A) Unit cell with NP radius $r=6$~nm, and height $h=30$~nm. Nearest neighbour spacing $l=24$~nm and lattice constant $a_0=65$~nm, (B) Alternating NPs are perturbed by $\delta r = \pm 0.5$~nm, with the height remaining constant, (C) BZ, (D, E) Spectral function: The highest energy band experiences polariton-like splitting. Drude losses $\gamma = 38$~meV.}}
    \label{fig:c6v_bulk}
\end{figure}

The existence of a Dirac cone is dependent on the crystal symmetry of the system. In plasmonics, the existing literature has focused on hexagonal, graphene-like arrangements of NPs \cite{han2009dirac, wang2016existence}, where deterministic Dirac cones appear at the $K$ and $K'$ points in the BZ and are protected by the $C_{3v}$ symmetry of these points \cite{lu2014dirac, makwana2018designing}. Extending beyond these simple lattices, we follow group theoretical rules for the existence of Dirac cones. Since the symmetries in the BZ are linked to the real space arrangement of NPs, these rules determine the lattices we can use. A more in depth discussion of the symmetries are given in Supplementary Material. We will consider three types of unit cells, with the symmetries shown in \autoref{fig:layout}A. In the following, we will refer to them by their pre-perturbation symmetry, $C_{3v}$, $C_{6v}$ and $C_{4v}$.

Whilst group theoretical concepts predict the presence of Dirac cones \cite{saba2017group}, the frequency at which they occur is dependent on the \rev{interaction matrix of the} system . Provided the cellular structure satisfies the required symmetries there \rev{exists a} freedom in the choice of other parameters such as the radius, number and type of NPs. Similarly, there is flexibility when breaking the Dirac cone. The \rev{degree of symmetry breaking by changes in the} orientation, size and material properties of these elements all influence the width of the band gap. In this regard, system parameters must be tuned to ensure that once the degeneracy is broken, a band gap across the whole BZ is achieved. \rev{Importantly,} the guidelines espoused herein are applicable to any nanophotonic or plasmonic system, such as arrangements of dielectric NPs or periodically doped graphene.

\autoref{fig:c3v_bulk}A,B shows the unit cell for the $C_{3v}$ case before, and after, the perturbation where the mirror symmetry $\sigma_v$ is removed by a rotation. \rev{In \autoref{fig:c3v_bulk}D,E we plot the band structures, including all neighbours in a QSA. Upon symmetry reduction, the Dirac cone is opened resulting in a complete band gap (blue).}
In \autoref{fig:c3v_bulk}F,G we characterise the system including retardation and radiative effects, and plot the spectral function \cite{zhen2008collective}. 
Since the NP and lattice parameters are very subwavelength, the QSA accurately describes the behaviour of the lower energy bands, although it does not capture the polariton splitting at the light line of the highest energy band. 
\rev{In these plots, we choose Drude losses $\gamma = 10~$meV for clarity, and results for material losses in silver, $\gamma = 38$~meV are given in the Supplementary Material.}
\rev{Three additional bands, at higher and lower energies, are also present in this system, since there are six elements in the unit cell. As they are not in the vicinity of the Dirac cone, we choose not to show them.}
A particular advantage of this lattice is that only a single type of NP is used, which makes fabrication more straightforward.

\rev{\autoref{fig:c6v_bulk}A,B shows the band structure for the $C_{6v}$ case, with $\gamma = 38$~meV. }For this case, both $\sigma_v$ and inversion symmetry must be broken to gap the Dirac cone. This is achieved by perturbing the radius of alternating NPs in the unit cell by $\pm\delta r$. \rev{The height of the NPs is fixed.} We plot the \rev{spectral function only since the quasistatic eigenvalue problem becomes non-linear as a result of employing spheroids of different sizes.}
Note that this case is similar to transition metal dichalcogenides and hexagonal boron nitride, which naturally have broken inversion symmetry unit cells due to their bipartite lattice structures with different atoms. 
\rev{Unlike the $C_{3v}$ case, here there are two Dirac cones at different frequencies which both result in a complete band gap once they are opened.} Previous plasmonic topological valley systems have solely focused on single band effects, but such a dual band structure allows topological valley states to exist in two frequency regimes. Dual band effects have also been demonstrated in photonic crystals \cite{chen2019valley} and acoustics \cite{zhang2019subwavelength}.

\rev{Dirac cones are only guaranteed for these triangular and hexagonal lattices. However, there are other instances where Dirac cones can be engineered along high-symmetry lines \cite{sakoda2014photonic,he2015emergence, makwana2019tunable}. Notably, irrespective of whether the Dirac cone is symmetry induced or not, there remains a non-trivial valley-Chern number defined in the vicinity of the degeneracy \cite{he2015emergence}. We give an example of a plasmonic metasurface possessing these properties in the Supplementary Material.}

\subsection{Edge states}\label{sec:edge_states}
\begin{figure}
    \centering
    \includegraphics[width=\columnwidth]{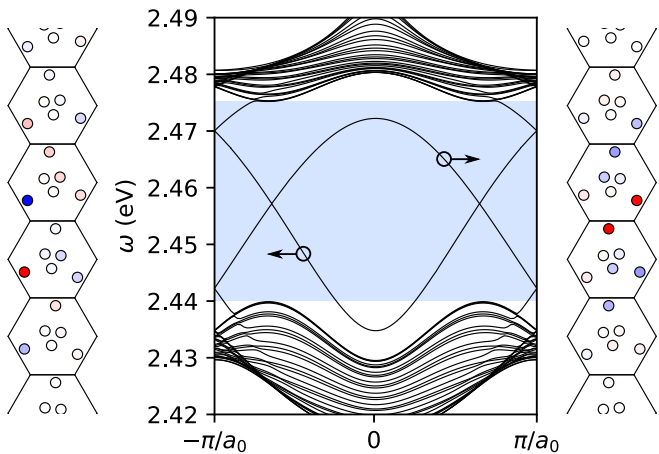}
    \caption{$C_{3v}$, Type I edge: The upper and lower media are perturbed by $\theta = \pm0.1$~rad. Band structure for the edge states. The dipole moments of the edge states circled are shown either side, demonstrating the confinement to the edge.}
    \label{fig:c3_type1_ribbon_dispersion}
\end{figure}
We now investigate edge states that reside in the band gaps generated in the previous subsection. The bulk boundary correspondence guarantees the existence of topological valley edge states between topologically distinct regions characterised by valley-Chern numbers with opposite sign \cite{qian2018topology, silveirinha2019proof}. Trivial edge states are also possible between regions with the same valley-Chern number. In \autoref{tab:edge_types}, we summarise the perturbations of two regions, A and B, which generate non-trivial and trivial edge states. There are, at most, three geometrically distinct interfaces possible for each of the triangular-based structures.
For each of these cases there will be two inequivalent interfaces A/B and B/A, that yield different eigenstates.

The existing literature in plasmonics on topological valley effects solely considers one of the three distinct interfaces that are possible \cite{jung2018midinfrared, wu2017direct}; specifically, the Type I case shown in \autoref{tab:edge_types}. Herein the realisation of the topological Type II edge states in the $C_{3v}$ system reveals features which have yet to be discussed in the context of plasmonics. 
Also, although trivial Type III edge states are not guaranteed to exist, they do here for this choice of parameters, (see Supplementary Material). Notably, the $C_{3v}$ and $C_{6v}$ Type I cases are mathematically identical to canonical honeycomb structure which has been the subject of other works \cite{chen2017valley, yang2018topological}, as they share the same symmetries.

\begin{table}
\begin{tabularx}{\columnwidth}{|c|c|c|X|}
\hline
Type & A & B & Properties \\
\hline
I   & $+\theta$ & $-\theta$ & Two edge states propagating \\
    &           &           &  in opposite directions\\
\hline
II  & $+\theta$ & $+\theta + \frac{\pi}{3}$ & Single edge state in band gap\\
\hline
III & $+\theta$ & $-\theta + \frac{\pi}{3}$ & Two nearly degenerate edge states\\
\hline
\end{tabularx}
\caption{$C_{3v}$ interfaces: Summary of the different interface types, including the perturbations of regions A and B and properties of each edge state.}
\label{tab:edge_types}
\end{table}

\begin{figure}
    \centering
    \includegraphics[width=\columnwidth]{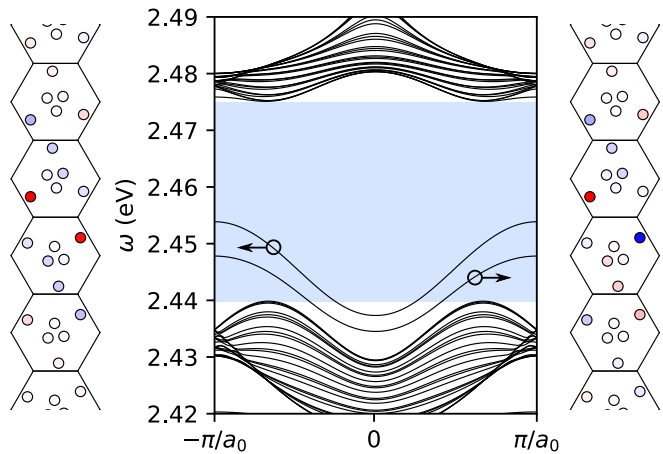}
    \caption{$C_{3v}$, Type II edge: The upper and lower media are perturbed by $\theta = +0.1$~rad and $\theta =+0.1+\frac{\pi}{3}$~rad, respectively. \rev{Only modes of one interface are present in the bulk band gap and the dipole moments of these modes are shown either side.}}
    \label{fig:c3_type2_ribbon_dispersion}
\end{figure}

To numerically simulate the edge states, we set up a ribbon of hexagons and set Bloch periodicity across the ribbon. The interactions between NPs include all neighbours in a QSA. Along the finite direction, we clad region A with region B on the top and bottom to capture both interfaces A/B and B/A. Each region is $N=10$ unit cells long to ensure that edge states at each interface decay within the length of the ribbon. At each end of the ribbon, we choose a hard boundary condition with the vacuum. Edge states which appear at this boundary are not topological (since the valley-Chern number of the vacuum is zero); these are removed from the band structures.

\autoref{fig:c3_type1_ribbon_dispersion} shows the band structure for $C_{3v}$, Type I edges. There are two overlapping edge states across the band gap which exist over the same frequency range. On either side of band structures we plot the dipole moments of the eigenstates of each interface; these two eigenstates show a clear difference in the pattern of the mode. Along both interfaces, the edge state is very subwavelength and strongly confined.  

The band structure for $C_{3v}$, Type II edges is shown in \autoref{fig:c3_type2_ribbon_dispersion}. Unlike \autoref{fig:c3_type1_ribbon_dispersion}, both edge states now reside on the same interface. They also do not span the band gap due to the size of the perturbation. 
Whilst the bands do not span the band gap, they can still be used to filter modes: Only A/B interface modes exist in the gap meaning there are no modes along B/A to couple into.

\begin{figure}
    \centering
    \includegraphics[width=\columnwidth]{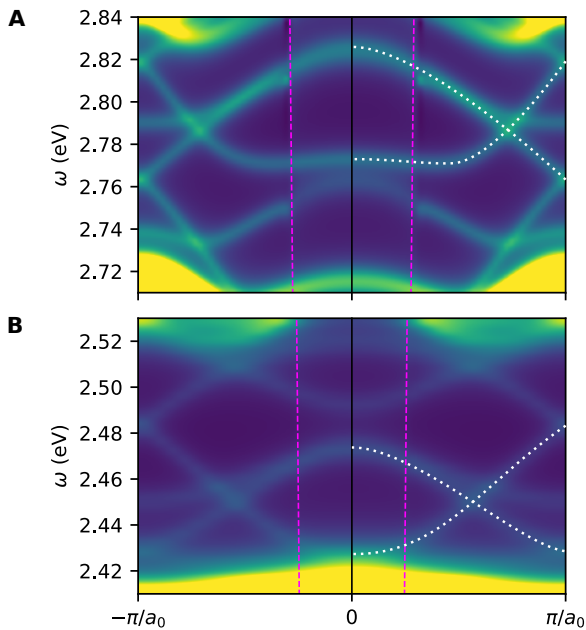}
    \caption{\rev{$C_{6v}$, Type I edge: The upper and lower media differ by an inversion about the vertical. Edge states reside in both (A) upper band gap and (b) lower band gap. Relevant topological edge states are highlighted on the right in white.}}
    \label{fig:c6v_ribbon}
\end{figure}

\rev{Finally, \autoref{fig:c6v_ribbon} shows the band structure for the edge modes in the $C_{6v}$ lattice. As with the bulk band structure, we plot the spectral function since the quasistatic eigenvalue problem is non-linear. The higher frequency band is shown in \autoref{fig:c6v_ribbon}A and the lower frequency band in \autoref{fig:c6v_ribbon}B. Since the spectral function captures all modes, including those which are at the boundary with the vacuum, we highlight in white the topological edge states on the relevant interfaces.  We see that both bands possess counter propagating edge states, which will be exploited in \autoref{sec:manipulating}.}

\begin{figure}
    \centering
    \includegraphics[width=\columnwidth]{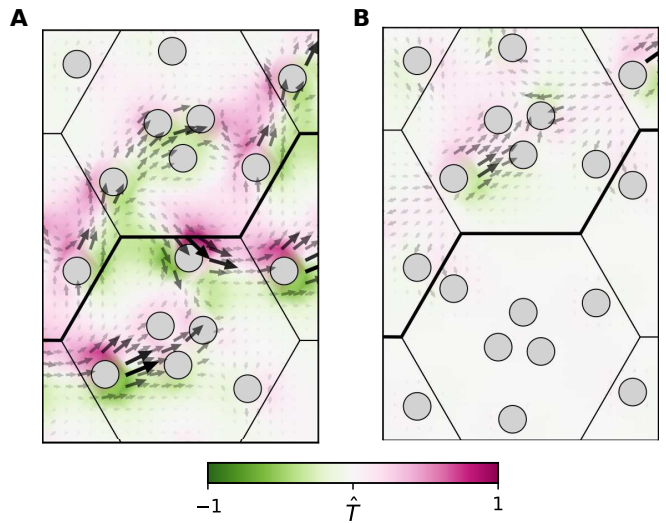}
    \caption{Chirality comparison: Poynting vector $\textbf{S}$ (arrows) and normalised spin angular momentum (colour) $\hat{T}_z$. (A) Non-trivial edge mode on $C_{3v}$ Type I edge. (B) Trivial edge mode, $C_{3v}$ Type III edge: Energy flow is more linear, as seen in the small spin angular momentum amplitude.}
    \label{fig:chirality_comparison}
\end{figure}

\subsection{\rev{Elements of topological protection}}
Topological insulators that break time-reversal symmetry have strictly unidirectional edge states. In contrast, the presence of time-reversal symmetry for our system results in bidirectional edge states, hence backscattering is not entirely prohibited. Despite this, there are still elements which contribute to these states having a degree of protection that allows them to be used to efficiently guide light.

A mode with a particular chirality will not readily couple with one of opposite chirality, giving it protection against disorder which does not cause the chirality to switch. 
We calculate quantities that characterise spin and chirality, to emphasise the difference between the topologically trivial and non-trivial edge states. The time averaged Poynting vector $\textbf{S} = \frac{1}{2}\text{Re}(\textbf{E}\times\textbf{H}^*)$ characterises the flow of electromagnetic energy in a system. In \autoref{fig:chirality_comparison}, we plot the Poynting vector (arrows) for a non-trivial, Type I edge state and a trivial, Type III edge state. In the non-trivial case we see chirality appears in the Poynting vector but in the trivial case it remains mostly linear. We also plot the out of plane component of the spin angular momentum $T_z = \text{Im}(\textbf{H}^*\times\textbf{H})_z$ in the plane of the metasurface (colour) since in dispersive, inhomogeneous media, this quantity provides a more appropriate description of chiral light matter interactions \cite{bliokh2017optical}. Again, in the non-trivial case, the normalised magnitude of the spin angular momentum is much greater than the trivial case. 

The presence of backscattering in topological valley states is also intimately linked to the separation of forwards and backwards propagating modes in Fourier space \cite{makwana2018designing}. In the $C_{3v}$ and $C_{6v}$ cases discussed in \autoref{sec:dirac_cones}, after breaking the necessary symmetries, valleys at $K/K'$ are well separated in Fourier space as they are at the edges of the BZ. 
\rev{By contrast, in systems with Dirac points engineered along high-symmetry lines, rather than high-symmetry points, such as the $C_{4v}$ system; valleys will not be maximally separated, see Supplementary Material}.
\rev{This results in these systems} being more susceptible to backscattering. Despite this there are means in which to expand the Fourier separation\rev{, the separation between opposite valleys in reciprocal space,} for this particular case \cite{makwana2019tunable}. \rev{Notably, armchair-type edges in triangular lattices also have a small Fourier separation relative to their zig-zag counterparts. However, unlike the $C_{4v}$ system, considered here, the valleys are completely coupled in the vicinity of $\Gamma$ resulting in an anti-crossing and enhanced backscattering \cite{bi2015role}.}

Other popular designs, such as the proposal from Ref. \cite{wu2015scheme}, also suffer from this; there an extended honeycomb lattice unit cell is used to emulate the quantum spin Hall effect. \rev{However, the larger unit cell results in the folding of the $K/K'$ valleys onto $\Gamma$ thereby inhibiting the protection afforded by a larger Fourier separation. Additionally, upon the introduction of an interface, the $C_6$ reliant pseudospin states couple to each other resulting in an anti-crossing near $\Gamma$.  These properties result in reduced protection against positional disorder and hence these modes are far less robust than the topological valley modes considered here \cite{orazbayev2019quantitative}.}

\section{Topological Valley Networks}\label{sec:manipulating}

Using the topological valley edge states, we can now design topological networks \cite{makwana2018designing}. These are formed by joining geometrically distinct regions of the various symmetry types. Again, we will begin with $C_{3v}$ systems before moving onto the $C_{4v}$ system. 

In the following, we include full electromagnetic interactions between NPs using the retarded, radiative Green's function. We plot the absolute value of real part of the dipole moment for each NP in the lattice. To make the edge states more visible, we increase the radius of the NPs in the plots but this does not change the physical system. In each case, we excite a particular NP in the respective lattice located near an interface which triggers a mode to propagate along the interface on which it sits. \rev{Initially}, the Drude losses are set to zero to fully expose the behaviour of the edge states. To prevent the mode reflecting off the hard boundary with the vacuum, we gradually increase the losses towards these boundaries. Since the topologically non-trivial edge states possess a chirality, a circularly polarised magnetic dipole source external to the lattice could also be used to excite unidirectional modes, where the direction is determined by the spin angular momentum at the source position \cite{proctor2019exciting}.

\begin{figure}
\centering
\includegraphics[width=\columnwidth]{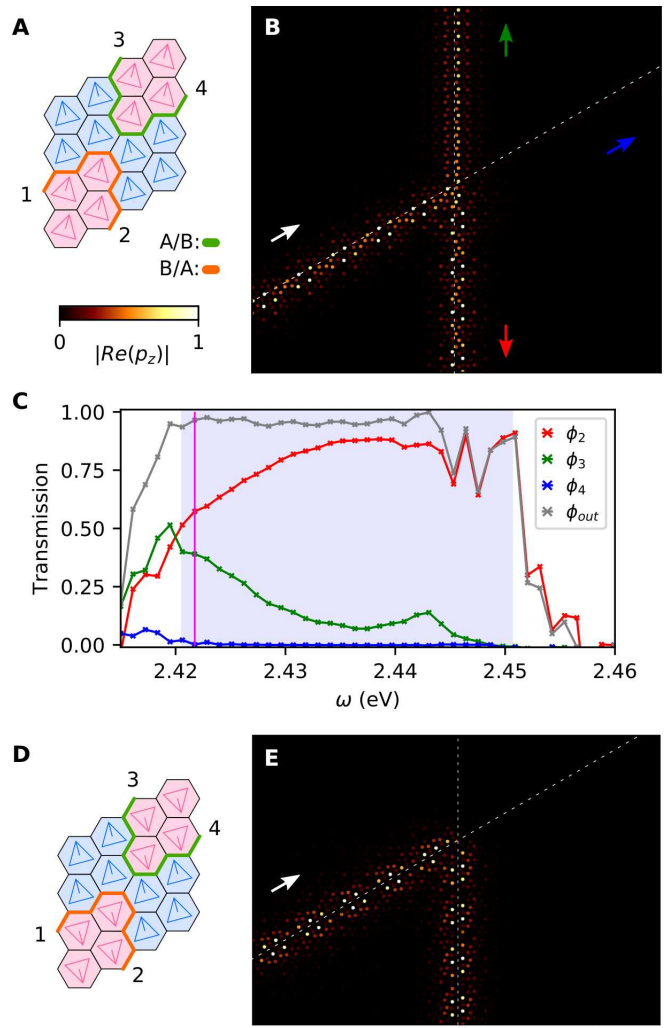}
\caption{
\rev{Topological waveguiding with $C_{3v}$ networks. 
(A) Type I: Two-way splitting junction, 
(B) $\omega=2.421$~eV. The mode along lead 1 splits at the junction, propagating around the gentle $2\pi/3$ (lead 3) and sharp $\pi/3$ (lead 2) bends, 
(C) Transmission $\phi$ along each lead. The band gap is highlighted in blue,
(D) Type II: Filtering junction, 
(E) $\omega=2.426$~eV. The mode only propagates around the sharp $\pi/3$ bend owing to there being no mode, at this frequency, along the other interface to couple into.}
}
\label{fig:c3v_scattering}
\end{figure}

A junction of the $C_{3v}$, Type I system is shown in \autoref{fig:c3v_scattering}A,B. The semi-analytical CDM gives a clarity to the dipole moment mode patterns along each lead, which correspond to the edge states calculated in \autoref{fig:c3_type1_ribbon_dispersion}. The patterns along lead 1 and lead 2 are identical, but they differ from lead 3, which highlights the different interfaces A/B and B/A as shown in \autoref{fig:c3v_scattering}A. The A/B mode is able to couple into the B/A mode at the junction due to the overlapping edge states. Despite this the mode does not propagate straight through along lead 4 due to it having opposite chirality and a wavevector mismatch \cite{makwana2018designing}. 

To highlight the robustness of this junction, we calculate the time-averaged Poynting flux through each of the exit leads \rev{$\phi_i$} as a fraction of the flux along the entry lead \rev{$\phi_1$}. In the frequency range where both interface modes are excitable, the majority of energy flows along lead 2. The blue region between lead 1 and 3 in \autoref{fig:c3v_scattering}A introduces a barrier to energy flow; similar effects have been observed in other systems \cite{wu2017direct} and the importance of the nodal region was highlighted in \cite{makwana2018designing}. \rev{Importantly, over the majority of the band gap, $\phi_\text{out}$, the fraction of output over input energy is $\sim$1. As shown in \autoref{fig:c3_type1_ribbon_dispersion}, the topological valley bands do not reach the top of the band gap causing $\phi_\text{out} < 1$ at higher frequencies. We note that the band gap is shifted $\sim 0.02$~eV compared to the quasistatic band structure due to the radiative correction to the polarisability.}

Filtering using the $C_{3v}$, Type II system is shown in \autoref{fig:c3v_scattering}D,E. When the mode along lead 1 reaches the junction it only propagates around the sharp $\pi/3$ bend, owing to the identical interfaces along lead 1 and lead 2; as seen in the identical mode patterns. As we showed in \autoref{fig:c3_type2_ribbon_dispersion}, there \rev{are only edge states} in the band gap corresponding to the B/A interface. Therefore, unlike the Type I edge, there is no mode to couple into along lead 3 at this frequency resulting in this filtering effect.

\begin{figure}
    \centering
    \includegraphics[width=\columnwidth]{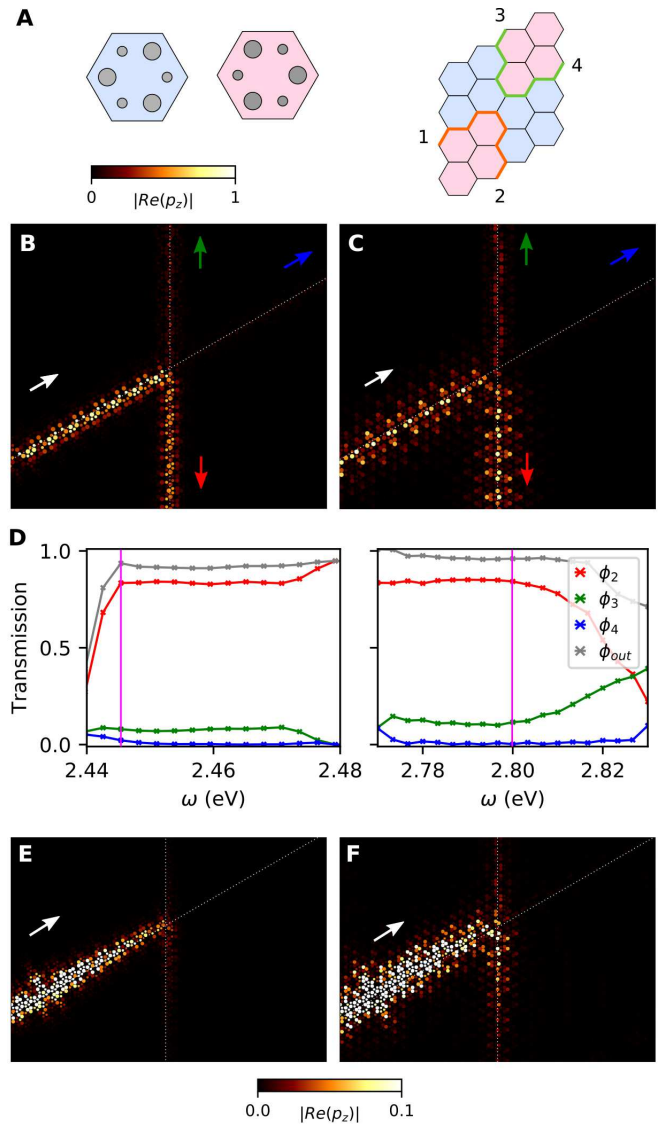}
    \caption{\rev{Dual-band waveguiding with $C_{6v}$ networks.
    (A) Diagram of the unit cells and two-way splitting junction,
    (B) $\omega=2.800$~eV. The mode propagates around both gentle and sharp bends, with almost zero propagation along lead 4.
    (C) $\omega = 2.446$~eV. The mode also propagates and splits at lower frequencies.
    (D) Transmission $\phi$ along each lead over the frequencies at which topological modes are excited, for lower and higher frequency bands, 
    (E, F) Modes excited at the same frequency as (A, B) including Drude losses $\gamma = 10$~meV.}
    }
    \label{fig:c6v_scattering}
\end{figure}

The previous two examples have a specific choice of nodal region at which the different regions intersect, which limits the splitting of a mode, in the Type I case, into two directions only. By carefully choosing the arrangement of this intersection point, \rev{it is possible} to realise \rev{topological} four-way splitting \rev{as well as trivial five-way splitting} of a mode \cite{makwana2018designing}. We show examples of both of these effects in the Supplementary Material.

\rev{We now move onto waveguiding with the $C_{6v}$ edge states. Networks are set up with the unit cell arrangements shown in \autoref{fig:c6v_scattering}A. In \autoref{fig:c6v_scattering}B a mode is excited along lead 1 in the upper band gap of the system. As with the $C_{3v}$, Type I edge states, the mode splits at the junction and the mode patterns confirm the distinct interfaces.  In \autoref{fig:c6v_scattering} we show that the lower band edge states also possess the same splitting properties. The Poynting flux through each of the leads as a fraction of the flux through the entry lead is shown in \autoref{fig:c6v_scattering}D, for the frequencies at which topological modes are excited. As with the $C_{3v}$ case, the fraction of output transmission over input is $\sim$ 1 for the topological modes. In addition, after studying in detail the behaviour of the topological valley state, we also show the effect of losses on an interface for the $C_{6v}$ case with $\gamma = 10~$meV, which cause the edge state propagation to become attenuated (see \autoref{fig:c6v_scattering}E, F)}

\section{Conclusion}

The existence of subwavelength, topological valley states in plasmonic metasurfaces has been presented. Using systematic symmetry arguments we provide a guide to realise topological valley modes in any nanophotonic system. Extending the existing topological nanophotonics literature, we describe all the possible topological valley modes available for hexagonal unit cell systems, which include new non-trivial and trivial edge states. Furthermore, by designing Dirac cones in square lattice systems we reveal \rev{emergent} three-way energy splitting in a plasmonic metasurface for the first time. The semi-analytical CDM used to model the systems in this work removes any computational complexity to expose the topological properties and behaviour of modes we engineer. Our results are however general and could be applied also to lower frequency plasmonics such as periodically doped graphene or graphene islands, or spoof surface plasmons. 

\begin{acknowledgments}
M.P., P.A.H., S.A.M. and R.V.C. acknowledge funding from the Leverhulme Trust. P.A.H. also acknowledges funding from Funda\c c\~ao para a Ci\^encia e a Tecnologia and Instituto de Telecomunica\c c\~oes under projects CEECIND/03866/2017 and UID/EEA/50008/2019. S.A.M., R.V.C. and M.P.M. thank the UK EPSRC for their support through Programme Grant EP/L024926/1.
\end{acknowledgments}

\clearpage
\bibliography{library}

\end{document}